\documentclass[12pt]{article}
\usepackage[latin1]{inputenc}
\usepackage{graphicx}

\usepackage{scicite}

\usepackage{times}
\usepackage{amsmath}

\newenvironment{sciabstract}{%
\begin{quote} \bf}
{\end{quote}}

\newcounter{lastnote}
\newenvironment{scilastnote}{%
\setcounter{lastnote}{\value{enumiv}}%
\addtocounter{lastnote}{+1}%
\begin{list}%
{\arabic{lastnote}.}
{\setlength{\leftmargin}{.22in}}
{\setlength{\labelsep}{.5em}}}
{\end{list}}

\title{Toward Forecasting Volcanic Eruptions using Seismic Noise}

\author
{Florent Brenguier,$^{1,2\ast}$ Nikolai M. Shapiro,$^{2}$ Michel Campillo,$^{1}$ Valérie Ferrazzini,$^{3}$\\ Zacharie Duputel,$^{3}$ Olivier Coutant,$^{1}$  and Alexandre Nercessian$^{2}$\\
\\
\normalsize{$^{1}$Laboratoire de Géophysique Interne et de Tectonophysique,}\\
\normalsize{CNRS \& Univ. Joseph Fourier, Grenoble, France}\\
\normalsize{$^{2}$Laboratoire de Sismologie, Institut de Physique du Globe de Paris,}\\
\normalsize{CNRS, Paris, France}\\
\normalsize{$^{3}$Observatoire Volcanologique du Piton de la Fournaise, IPGP,}\\
\normalsize{La Réunion, France}\\
\\
\normalsize{$^\ast$To whom correspondence should be addressed; E-mail:  florent.brenguier@ujf-grenoble.fr.}
}

\date{}

\begin{document} 


\baselineskip24pt


\maketitle


\begin{sciabstract}
During inter-eruption periods, magma pressurization yields subtle changes of the elastic properties of volcanic edifices. We use the reproducibility properties of  the ambient seismic noise recorded on the Piton de la Fournaise volcano to measure relative seismic velocity variations of less than $\textbf{0.1 \%}$  with a temporal resolution of one day. Our results show that five studied volcanic eruptions were preceded by clearly detectable seismic velocity decreases within the zone of magma injection. These precursors reflect the edifice dilatation induced by magma pressurization and can be useful indicators to improve the  forecasting of volcanic eruptions.
\end{sciabstract}

Volcanoes are among the most dynamic geological objects and their eruptions provide a dramatic manifestation of the Earth's internal activity. However, strong eruptions are only short episodes in the history of volcanoes which remain quiescent most of time. During this inter-eruption periods, slow 
processes such as changes in the magma supply to the reservoir and changes of the magma 
physical or chemical properties lead to perturbations of the reservoir pressure and to preparation 
of new eruptions. It is, therefore, very important to better describe these processes to fully understand 
the functioning of active magmatic systems  and to improve our ability to forecast volcanic eruptions. 

The inter-eruptive magma pressurization  and/or intrusions of dykes result in subtle changes in the shape of the volcanic edifices that may be detected by
modern strainmeters and tiltmeters \cite{peltier05} or by satellites \cite{Massonnet1995}. The main limitation of these geodetic methods is that they are based on the interpretation of displacements or tilts observed near the surface, which limits their sensitivity to changes located at depth.
Volcanic seismicity provides informations about  short-term (few seconds to few days) mechanical processes occurring within volcanoes at depth. The spatio-temporal evolution of magma migration can be retrieved by precise location of seismic events \cite{battaglia05} and by accurate determination of seismic source properties \cite{Chouet96,Chouet2003}.  However, analysis of volcanic seismicity can not be used to detect the aseismic magma pressurization preceding fracturing and magma migration.
The inter-eruptive processes also produce perturbations of the elastic properties of volcanic edifices. They can be detected as changes in the travel times of seismic waves propagating within the volcanic edifices by using coda waves from repetitive seismic sources such as multiplets \cite{poupinet84,Ratdomopurbo95, wegler06} or small summital volcanic explosions  \cite{snieder04,gretMerapi} or correlations of a diffuse seismic wavefield excited by long-period seismo-volcanic events \cite{Sabra06}. However, these methods based on seismo-volcanic sources do not provide information about time periods when volcanoes are seismically quiescent. More recently, repeated seismic tomographies \cite{D.Patane08112006} has been used to
detect seismic velocity variations within Mount Etna over periods of a few years. This method reveals variations of the internal volcanic structure before and after an eruption. However, the construction of repetitive tomographic images requires long time periods of observation of seismicity and cannot
be done continuously. Moreover, the accuracy of repetitive tomographies is limited and may be insufficient to detect small velocity variations (less than 1 $\%$) associated with magma pressurization processes.

In this paper, we go beyond the limitations of methods based on seismo-volcanic sources by recovering temporal velocity variations within the Piton de la Fournaise volcano using continuous ambient seismic noise records. 
The basic idea is that a cross-correlation of random seismic wavefields such as coda or noise recorded at two receivers yields the Green function, i.e., the impulse response of the medium at one receiver as if there was a source at the other \cite{weaver01,campillopaul03,campillo2006,wapenaar06}. 
This property has been used for imaging the crust at regional scales \cite{shapiro05,Sabra05,yang07} and, more recently, has been applied to infer the internal structure of the Piton de la Fournaise volcano at La Réunion island \cite{brenguierPiton}. 
By computing noise cross-correlations between different receiver pairs for consecutive time periods, we make each receiver to act as a virtual highly repetitive seismic source. The associated reconstructed seismic waves (Green functions) can then be used to detect temporal perturbations associated to small velocity changes (less than 1 $\%$), \cite{stehlyInstru}.

\begin{figure}[htbp]
	\centering
		\includegraphics[width=0.9\textwidth]{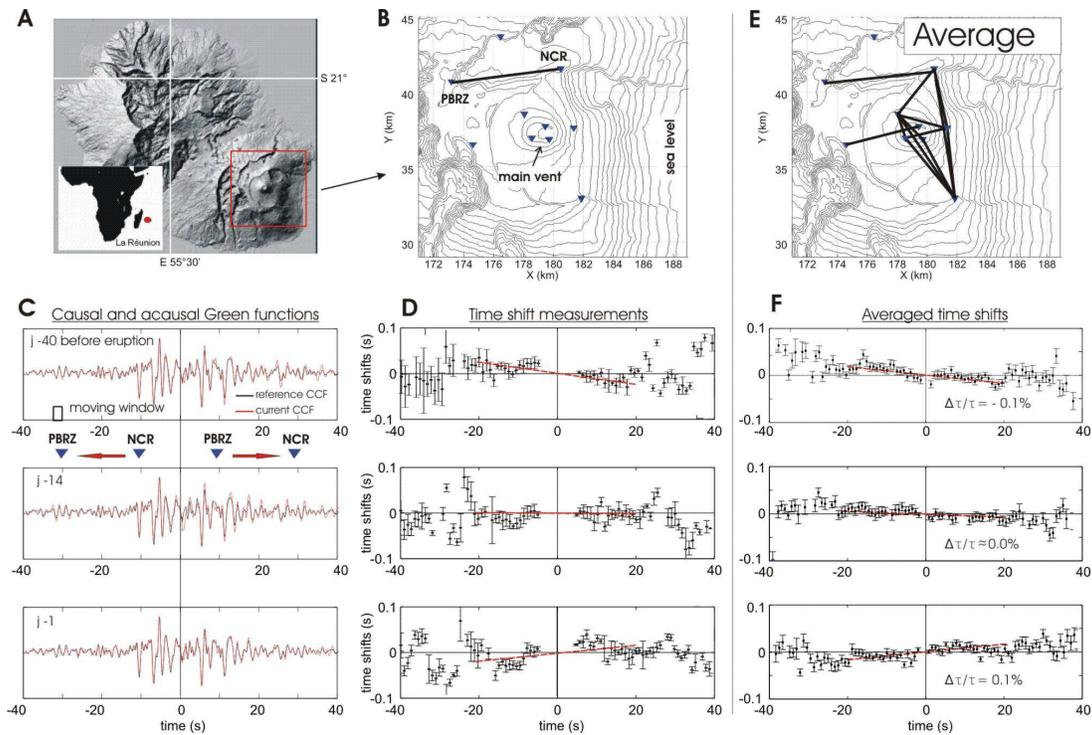}
	\caption{Measurements of relative velocity perturbations ($\Delta v/v$). \textbf{(A)} La Réunion Island. \textbf{(B)} Topographic map showing a direct path between two seismic receivers (PBRZ-NCR) indicated with inverted triangles. \textbf{(C)} Comparison  between reference (black curves) and current  (red curves) cross-correlation functions computed 40 days, 14 days, and one day prior to the eruption of June 2000 (day 359). \textbf{(D)} Time shifts measured between the reference and current cross-correlation functions computed between PBRZ and NCR. Red lines show results of the linear regressions. \textbf{(E)} Topographic map showing 13 receiver pairs used for averaging the time shift measurements. \textbf{(F)} Averaged time shifts. Resulting measurements of relative time perturbations (RTP, $\Delta \tau/\tau=-\Delta v/v$) estimated by linear regressions are illustrated with red lines.}
	\label{fig:FigureMesDt}
\end{figure}

We applied this method to study the Piton de la Fournaise volcano on La Réunion island (Fig. \ref{fig:FigureMesDt}A). 
%This basaltic volcano is one of the most active volcanoes in the world. 
During the past two centuries the average time between consecutive eruptions of this volcano has been about 10 months \cite{Stieltjes1989}. Geodetic and seismic data suggest that the Piton de la Fournaise eruptions are triggered by magma over-pressure within a magma reservoir located below the main vent (Fig. \ref{fig:FigureMesDt}B) at approximately sea level \cite{Nercessian96,peltier05}.  
%The so-called recorded seismic noise is mainly composed of seismic waves continuously generated by the ocean-coast interaction. 
We used the continuous seismic noise recorded between July 1999 and December 2000 by 21 vertical short period receivers operated by the Observatoire Volcanologique du Piton de la Fournaise to compute 210 cross-correlation functions corresponding to all possible  receiver pairs [see Fig. S1 and \cite{brenguierPiton} for details]. We used the spectral band between 0.1 and 0.9 Hz where the recovered Green functions has been demonstrated to consist of Rayleigh waves that are sensitive to the structure at depths down to 2 km below the edifice surface. The cross-correlation functions obtained by correlating 18 months of seismic noise are called the reference Green functions. The temporal evolution was then tracked by comparing the reference Green functions with current Green functions computed by correlating the noise from a ten-days-long moving window.
 
If the medium exhibits a spatially homogeneous relative velocity change $\Delta v/v$, the relative travel-time shifts ($\Delta \tau$) between the perturbed and reference Green functions is independent of the lapse time ($\tau$) at which it is measured and $\Delta v/v = -\Delta \tau/\tau = const$. Therefore, when computing a local time shift $\Delta \tau$ between the reference and the current cross-correlations in a short window
centered at time $\tau$, we would expect to find $\Delta \tau$ to be a linear function of $\tau$ (Fig. S2). By measuring the  slope of the  travel time shifts $\Delta \tau$ as function of time $\tau$, we finally estimate the Relative Time Perturbation (RTP) that is the opposite value of the medium's uniform relative velocity change ($\Delta v/v$). Fig. \ref{fig:FigureMesDt}C shows the reference and the current Green functions for one pair of receivers  (PBRZ-NCR) computed 40 days , 14 days, and one day before the eruption of June 2000 (day 359). Fig. \ref{fig:FigureMesDt}D shows the respective travel time shifts measured in the frequency band
[0.1-0.9] Hz using the Moving Window Cross Spectrum technique \cite{Ratdomopurbo95,supportingmaterial}. It can be seen in these measurements obtained just with one pair of receivers that the RTP changes from a negative to a nearly zero and then to a positive value 40 days, 14 days, and one day prior to the eruption, respectively. However, the RTP can be measured only very approximately when using only one pair of receivers. The accuracy  of the linear trend measurements is significantly improved by averaging local time shifts for different receiver pairs  assuming that the seismic velocities are perturbed uniformly within the sampled medium. We selected 13 receiver pairs  located near the caldera and showing good quality measurements  (Fig. \ref{fig:FigureMesDt}E)  to compute the averaged time shifts (Fig. \ref{fig:FigureMesDt}F) and to obtain accurate RTP estimates.

\begin{figure}[htbp]
	\centering
		\includegraphics[width=0.8\textwidth]{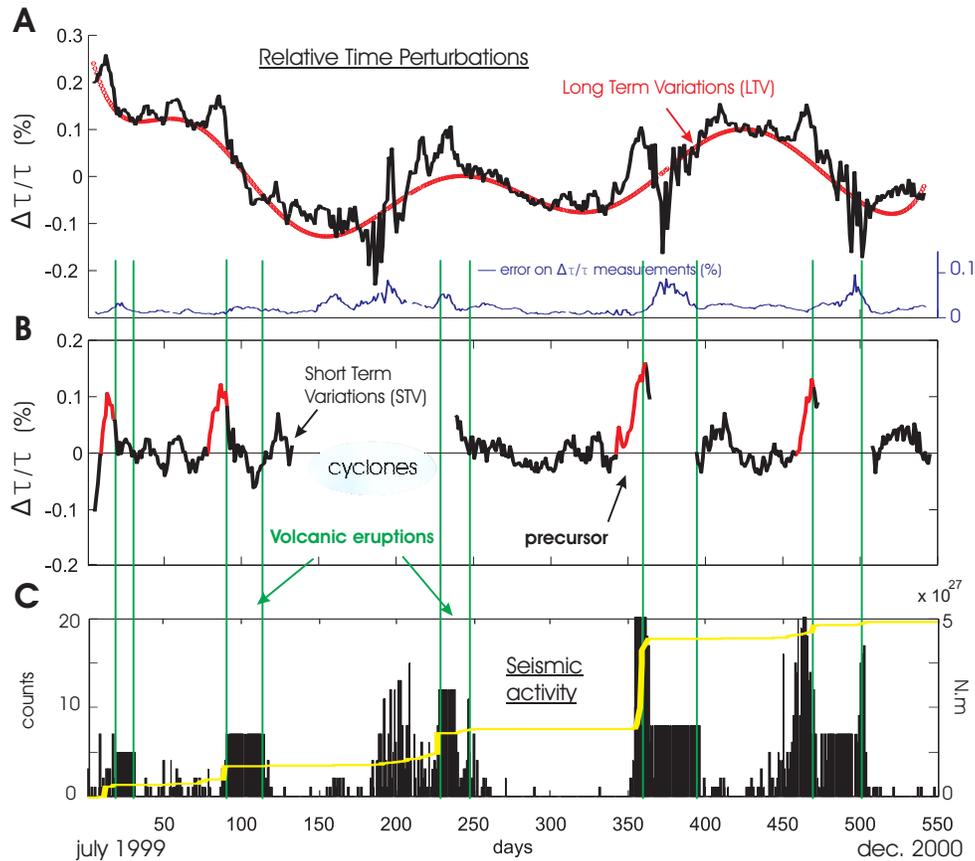}
	\caption{Evolution of the relative time perturbations (RTP) on the Piton de la Fournaise over 18 months. \textbf{(A)} Raw RTP over 18 months. The blue error curve represents the uncertainty of the linear slope estimation \cite{supportingmaterial}. Measurements with uncertainties higher than 0.04 $\%$ are excluded from the analysis. \textbf{(B)} Short-term variations (STV), RTP corrected from the long-term component (LTV) estimated as a polynomial function.  \textbf{(C)} Seismic activity, counted as a number of seismic events per day. The yellow line represents the cumulated seismic moment (scale on the right).}
	\label{fig:results}
\end{figure}

Fig. \ref{fig:results}A presents the relative time perturbations ($\Delta \tau/\tau$) measured over 18 months (June 1999 to December 2000). 
Intervals with low-quality measurements that correspond to periods of intense seismo-volcanic activity (Fig. \ref{fig:results}C) and to strong tropical cyclones (days 150-200) are excluded from the analysis (Fig. \ref{fig:results}B). The remaining time series show that the volcano interior exhibits changes at different time scales varying from a few months to a few days.
To simplify the analysis, we separate the short- and the long term-variations. We fit the Long Term Variations (LTV) by a polynomial function and then subtract it from the raw RTP to obtain the Short-Term Variations (STV, Fig. \ref{fig:results}B). 
The STV curve shows clear precursors to the volcanic eruptions characterized by an increase in the RTP and thus a decrease of seismic velocities. These precursors start about 20 days before the eruptions and correspond to relative velocity perturbations as small as 0.1 $\%$. 

We interpret the observed decreases in seismic velocities as an effect of the dilatation of a part of the edifice resulting from the magma pressurization within the volcano plumbing system similar to observations at Mount Etna \cite{DomenicoPatane03282003}. This interpretation is also supported by the clear opening of an individual fracture detected by an extensometer \cite{peltier06} prior to the  eruption n°4 (day 360, June 2000) synchronously with the RTP precursor (Fig. S3A). The lack of clear fracture opening precursors for the other eruptions would suggest that the detected dilatation may occur in a restricted zone and have non-measurable effects at the extensometer location.
The relative velocity changes return to a nearly background level during the eruption periods (Fig. \ref{fig:results}B). This observation is consistent with deflation associated to depressurization of the magma during its extrusion to the surface only few minutes after the beginning of the eruptions \cite{peltier06}.

\begin{figure}[htbp]
	\centering
		\includegraphics[width=0.6\textwidth]{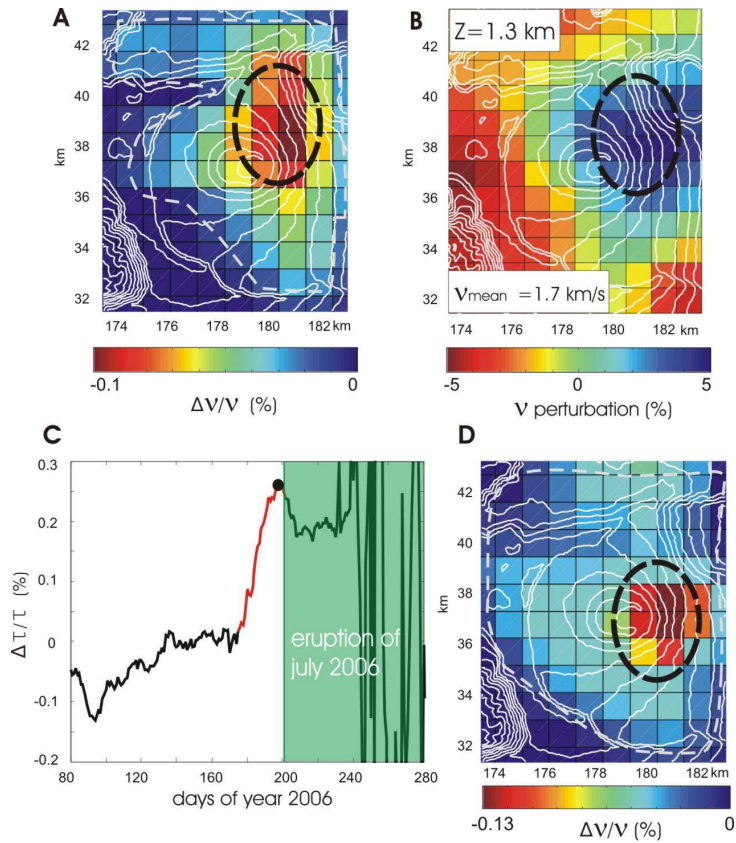}
	\caption{\textbf{(A)} Regionalization of the relative velocity perturbations associated with the second eruption precursor (day 85, Sept. 1999). The gray dashed line represents the limits of ray coverage. \textbf{(B)} Fast shear velocity anomaly imaged by passive surface wave tomography \cite{brenguierPiton} and interpreted as an effect of solidified dykes associated with the zone of magma injection. \textbf{(C)} Relative time perturbations before the eruption of July 2006. \textbf{(D)} Associated regionalized relative velocity perturbations one day before the beginning of the eruption.}
	\label{fig:CmpVs_dVs}
\end{figure}

The presented RTP curve required to spatially average the measured time shifts for different receiver pairs located in the vicinity of the main caldera. The same procedure applied to receiver pairs located outside the main caldera do not show eruption precursors. This suggests that the velocity variations are spatially localized. We apply a \emph{regionalization} procedure \cite{supportingmaterial} in order to localize more precisely the observed perturbations. We estimate the relative velocity variations for 28 individual receiver pairs selected according to quality criteria, subtract the long-term component from the raw measurements to obtain the short-term variations (STV) and then compute, in every grid cell, an average value from its neighborhood receiver pair paths.
As a result, we obtain for every day a two-dimensional map of relative velocity perturbations. A full set of this maps is shown in Movie S1 and a snapshot taken five days before eruption n°2 is presented in Fig. \ref{fig:CmpVs_dVs}A. The movie and  snapshot show that the precursors are not  distributed homogeneously in space but are mainly located in an area a few kilometers East from the main vent. 
%In term of methodology, whereas the spatial resolution is low (heterogeneous distribution of receiver pairs), this location does not correspond to the best resolved area (Fig. S4) and is thus probably not an artifact. 
This location nearly coincides with the fast velocity anomaly imaged by surface wave tomography at 1.3 km above sea level (Fig. \ref{fig:CmpVs_dVs}B), \cite{brenguierPiton}, and interpreted as an effect of solidified dykes associated to the zone of magma injection. This coincidence is an additional argument suggesting that the observed short-term seismic velocity decreases are produced by a dilatation associated with the pressurization within the volcano plumbing system.

The presented measurements are based on Rayleigh waves that dominate the noise cross-correlations. Therefore, the observed travel time variations mainly correspond to perturbations of shear wave velocity $\beta$. We use empirical laws for porous media \cite{pride03} to link the relative shear wave velocity perturbations to relative perturbations of porosity $\phi$ and volume $V$:

\begin{equation}
-\Delta \tau/\tau = \Delta \beta/\beta=-\frac{1}{2}\times  \Delta \phi/\phi=-\frac{1}{2}\times  \frac{\Delta V }{V\times\phi}
\end{equation}

\noindent
that can be, in turn, related to the over-pressure $\Delta P$ induced in the medium by the magma using a model of dilatancy:

\begin{equation}
\Delta V/V=\Delta P/K
\end{equation}

\noindent
where K is the incompressibility factor of the media that can be estimated as approximately 10 GPa by considering an average P-wave speed of 3400 $m/s$ \cite{brenguierPiton}, a density of 2000 $kg/m^{3}$, and a shear modulus value of 10 GPa.  We then take $\Delta \beta/\beta=-1\times  10^{-3}$ and an average realistic porosity of 0.1 and find $\Delta P$ to be approximately equal to 2 MPa or 20 bar. This over-pressure level is consistent with theoretical and experimental models of dike migration \cite{tait89,tait99}, with recent interpretation of strain data on the Soufrière Hills volcano of Montserrat \cite{voight06} as well as with the model of dyke propagation developed to explain the geodetic observations on the Piton de la Fournaise \cite{peltier06}.

The analysis of the long-term velocity variations show that they are associated to processes varying over several months (Fig. \ref{fig:results}A) with a possible seasonality. A clear seasonal dependance of the seismic velocity perturbations was observed from the analysis  of short-period noise cross-correlations at the Merapi volcano \cite{SensMerapi} and was related to variations of the depth of the superficial ground water layer because of precipitation. In the case of the Piton de la Fournaise, the observed LTV are not correlated with the pluviometric record on \emph{La Réunion} island (Fig. S3B). Moreover, the LTV amplitude and location (east of main vent) are very similar to what was observed for the short-term precursors (Fig. \ref{fig:results}B, S4E and Movie S2) suggesting that the long-term velocity variations reported in this study are related to the dynamics of the volcano-magmatic system.

Our results demonstrated that relative seismic velocity changes can be measured with an accuracy lower than 0.1\% by following the time evolution of the waveforms extracted from correlations of seismic noise. Application of this method to the data of the seismic network on the Piton de la Fournaise volcano revealed the evolution of the elastic properties of the upper volcano edifice during inter-eruption periods between July 1999 and December 2000. We detected systematic velocity decreases of the volcanic edifice preceding volcanic eruptions. These precursors clearly appeared approximately 20 days before four eruptions that occurred during the studied period. Approximate lateral positions of the maxima of observed velocity perturbations and a correlation with available geodetic data strongly indicate that the velocity variations are caused by dilatation of a part of the edifice. This process is caused by the pre-eruptive magma pressurization.

As a pilot experiment, we began to monitor the Piton de la Fournaise volcano by computing in real time the noise cross-correlations and measuring the velocity variations since spring 2006. This allowed us to identify a clear precursor for the last eruption that occurred in July 2006 (Fig. \ref{fig:CmpVs_dVs}C, D). Observation of such precursors may have important implications for monitoring volcano unrest and for improving the forecasting of volcanic eruptions. Detection of changes in elastic properties based on correlations of seismic noise may also be useful in other geophysical, engineering, and geotechnical applications that require non destructive monitoring of the media.

%%%%%%%%%%%%%%%%%%%%%%%%

%%%%%%%%%%%%%%%%%

%%%%%%%%%%%%%%%%%%
%%%%%%%%%%%%%%%%%

\newpage
\bibliography{ArtScience}

\begin{thebibliography}{10}
\bibitem{supportingmaterial}
Materials, methods are available as supporting material~on Science~Online .
\end{thebibliography}

\bibliographystyle{Science}

\begin{scilastnote}
\item 
All the data used in this work were collected by the seismological network of the Observatoire Volcanologique du Piton de la Fournaise. We are grateful to the Observatory staff. We thank L. Stehly, P. Gouédard, P. Roux,  L. de Barros and C. Sens-Schönfelder for helpful discussions. We are grateful to  A. Peltier and Meteo France for providing us with respectively extensometer and meteorological data. We thank F. Renard, I. Manighetti, and G. Poupinet for constructive comments concerning the manuscript. This work was supported by ANR (France) under contracts 05-CATT-010-01 (PRECORSIS) and COHERSIS.
\end{scilastnote}

\end{document}